\documentclass[a4paper,11pt]{article}
\usepackage{pos}
\usepackage[english]{babel}
\usepackage{amsmath}
\usepackage{float}
\usepackage{graphicx}
\usepackage{subcaption}
\usepackage{physics}
\usepackage{siunitx}
\usepackage{sidecap}
\usepackage{wrapfig}
\usepackage{dsfont}
\usepackage{color}
\usepackage{hyperref}
\usepackage{soul}

\newcommand{\ompe}{\omega_\mathrm{pe}}

\newcommand{\omci}{\Omega_\mathrm{i}}

\newcommand{\lse}{\lambda_\mathrm{se}}

\renewcommand{\printHeadAuthors}{Bohdan et al.}

\title{Using PIC and PIC-MHD to investigate cosmic ray acceleration in mildly relativistic shocks}
 \ShortTitle{Cosmic ray acceleration in mildly relativistic shocks}

\author[a, b]{Artem Bohdan}
\author*[c,d]{Anabella Araudo}
\author[c]{Allard Jan van Marle}
\author[e]{Fabien Casse}
\author[d]{Alexandre Marcowith}

\affiliation[a]{Max-Planck-Institut für Plasmaphysik, Boltzmannstr. 2, DE-85748 Garching, Germany}

\affiliation[b]{Excellence Cluster ORIGINS, Boltzmannstr. 2, DE-85748 Garching, Germany}

\affiliation[c]{Extreme Light Infrastructure ERIC, ELI Beamlines Facility, Za Radnici 835, CZ-25241 Dolni Brezany, Czech Republic }

\affiliation[d]{Laboratoire Univers et Particules de Montpellier (LUPM) Universit\'e Montpellier, CNRS/IN2P3, CC72, Place Eug\'ene Bataillon, F-34095 Montpellier Cedex5, France}

\affiliation[e]{Universit\'e de Paris, AstroParticle \& Cosmologie, CNRS CEA, Observatoire de Paris, PSL Research University, CNES, F-75013 Paris, France}

\emailAdd{artem.bohdan@ipp.mpg.de}

\abstract{Astrophysical shocks create cosmic rays by accelerating charged particles to relativistic speeds. However, the relative contribution of various types of shocks to the cosmic ray spectrum is still the subject of ongoing debate. Numerical studies have shown that in the non-relativistic regime, oblique shocks are capable of accelerating cosmic rays, depending on the Alfv\'enic Mach number of the shock. We now seek to extend this study into the mildly relativistic regime. In this case, dependence of the ion reflection rate on the shock obliquity is different compared to the nonrelativistic regime. Faster relativistic shocks are perpendicular for the majority of shock obliquity angles therefore their ability to initialize efficient DSA is limited. We define the ion injection rate using fully kinetic PIC simulation where we follow the formation of the shock and determine the fraction of ions that gets involved into formation of the shock precursor in the mildly relativistic regime covering a Lorentz factor range from 1 to 3. Then, with this result, we use a combined PIC-MHD method to model the large-scale evolution of the shock with the ion injection recipe dependent on the local shock obliquity. This methodology accounts for the influence of the self-generated or pre-existing upstream turbulence on the shock obliquity which allows study substantially larger and longer simulations compared to classical hybrid techniques.}

\ConferenceLogo{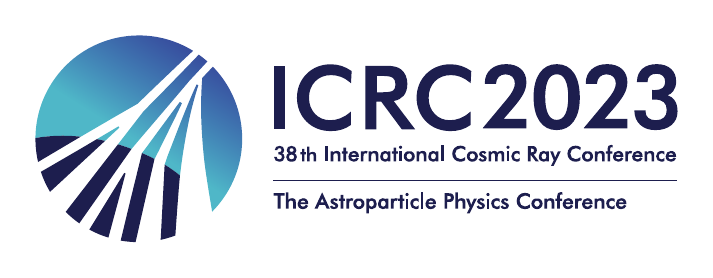}

\FullConference{%
38th International Cosmic Ray Conference (ICRC2023)\\
  26 July - 3 August, 2023\\
  Nagoya, Japan}

\begin{document}
\maketitle

\section{Introduction}

The origin and acceleration mechanisms of cosmic rays have been the subject of intense research in astrophysics. It is widely accepted that the bulk of galactic cosmic rays are produced in the aftermath of supernova explosions, where particles gain energy through Diffusive Shock Acceleration (DSA) \cite{Drury83}. However, the origin of Ultra-High-Energy Cosmic Rays (UHECRs) remains an enigma, and the quest to understand their sources continues to captivate the scientific community. One intriguing candidate for accelerating particles to extreme energies is found in the mildly relativistic shocks within the backflows of Active Galactic Nuclei jets \cite{Bell2019}. These shocks have the potential to accelerate particles up to energies of $10^{18}$ eV and beyond. 

In this study, we focus on mildly relativistic shocks with Lorentz factors ranging from 1 to 3, exploring their potential as sources of high-energy particles. The efficiency of DSA in these shocks depends heavily on their specific parameters, making them intriguing candidates for understanding the origin of UHECRs. By delving into the intricate details of particle acceleration within these shocks, we aim to shed light on the mechanisms responsible for producing cosmic rays with such extreme energies.

To gain insight into the acceleration processes, we need to combine two different methods, similarly how it was done for nonrelativistic oblique supernova remnant shocks \cite{vanMarle2022}. Firstly, we need to use the particle-in-cell (PIC) method to model the shock structure in order to determine the fraction of particles that gets reflected into the upstream medium. Once this has been established, we can use this injection rate to run further simulations with the PIC-magnetohydrodynamic (MHD) method to determine whether the injection rate suffices to trigger the {\bf non-resonant} streaming instability \cite{Bell2004,Bell2005}.

\section{PIC simulations}

\begin{figure}
    \centering
    \includegraphics[width=0.49\linewidth]{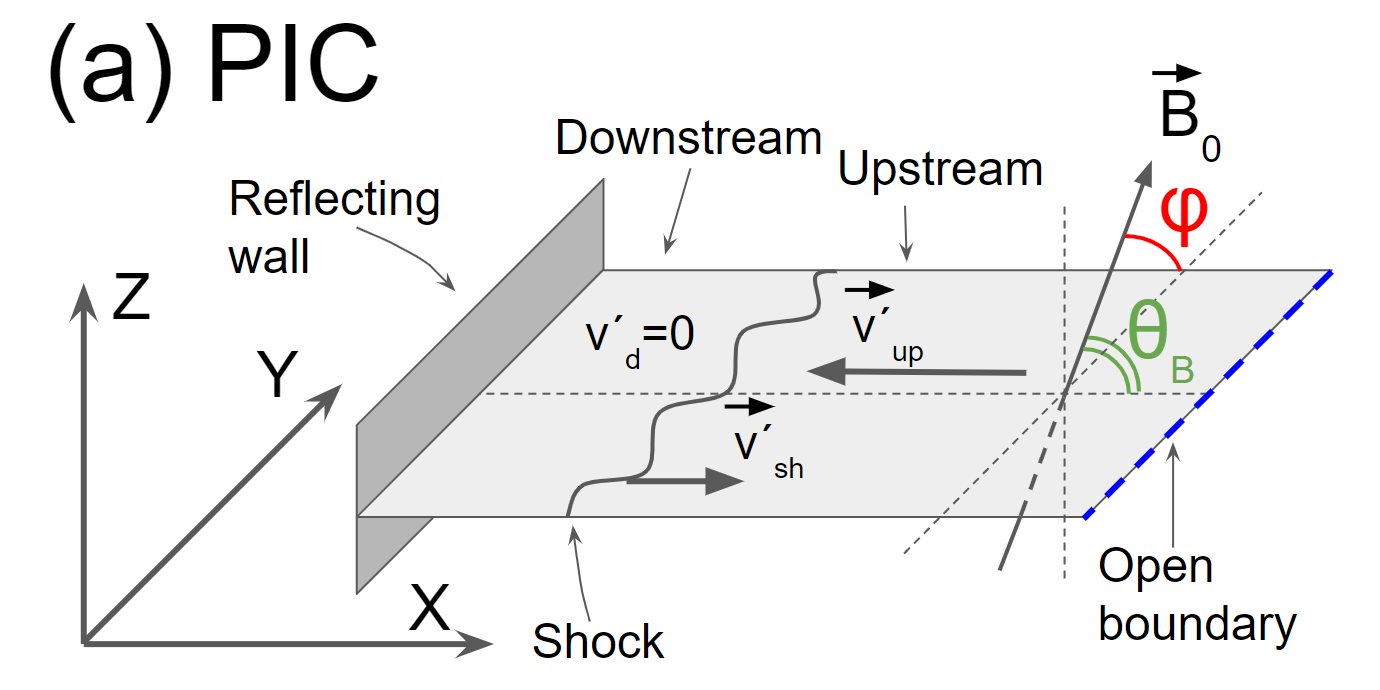}
    \includegraphics[width=0.49\linewidth]{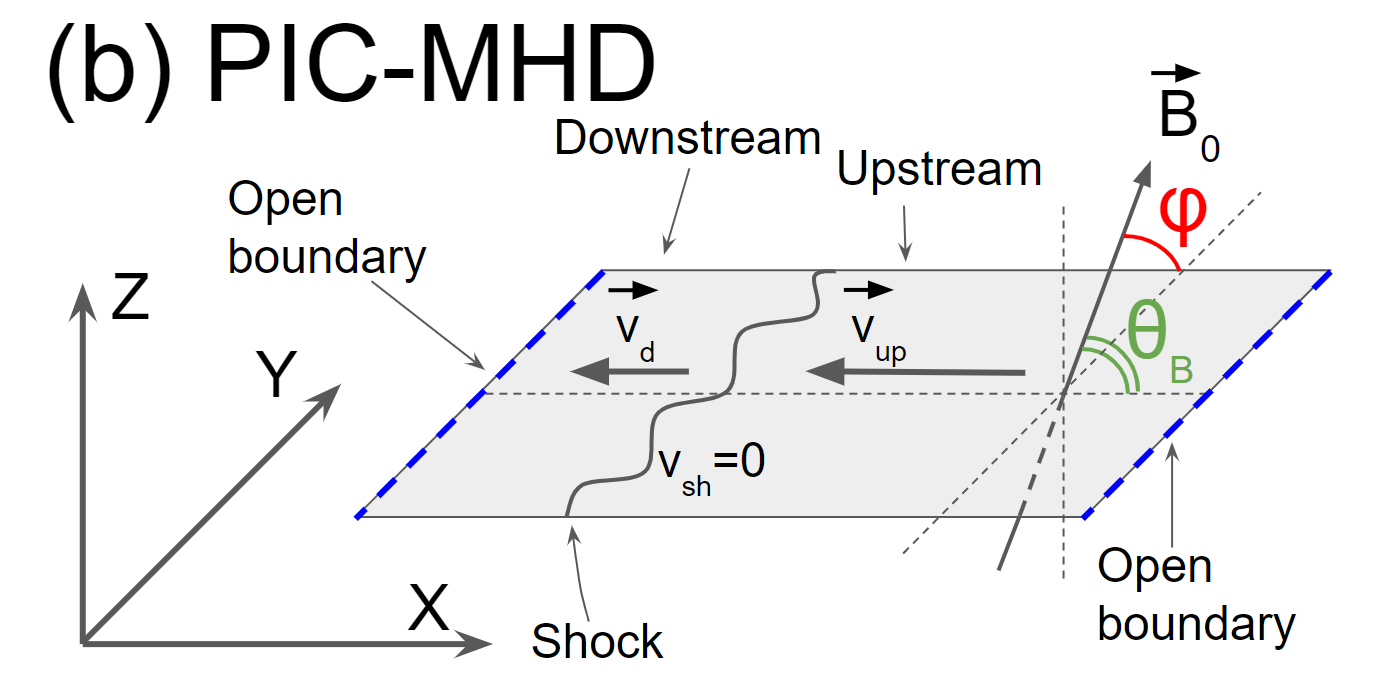}
    \caption{Simulation setups.}
    \label{fig:setup}
\end{figure}

PIC shock simulations are performed using an optimized fully-relativistic electromagnetic 2D PIC code with MPI-openMP hybrid parallelization developed from the TRISTAN code \citep{Buneman1993,Niemiec2008}.
Shocks are initialized using the reflecting-wall setup (Fig.~\ref{fig:setup}(a)). Interaction of the upstream electron-ion plasma flow with the reflecting wall results in a shock which propagates towards the upstream plasma flow in positive $x$ direction. The large-scale magnetic field, $\bf B_0$, is carried by the upstream plasma. The shock obliquity angle, $\theta_{B}$, is defined as the angle between $\bf B_0$ and the shock normal vector. Here we utilise the \emph{out-of-plane} magnetic field configuration, therefore $\varphi=90\deg$, where $\varphi$ the angle between the upstream magnetic field and y-axis which lies within the simulation plane. The upstream plasma velocity is chosen to have the resulting Lorentz factor of the shock of $\gamma_{\rm sh}=1.86$. This sets the Alfv\'enic and sonic Mach numbers as $M_{\rm A} \approx 20$ and $M_{\rm s} \approx 22$, which are defined in the upstream reference frame. The plasma beta value, which denotes the thermal-to-magnetic energy density ratio in the upstream region is $\beta =  1$. The ion-to-electron mass ratio is set to $m_i/m_e = 25$. To ensure adequate resolution of all important timescales in our simulation, we use a simulation timestep of $\delta t=1/16~\ompe^{-1}$ and a resolution of $\lse = c/\ompe = 8\Delta$, where $\Delta$ represents the size of one cell in our simulation grid. The number of particles per cell per species is $N_\mathrm{ppc} =40$.

We conducted a comprehensive study with 6 {\bf PIC} simulations, varying the obliquity angle $\theta_{B}$, as presented in Table~\ref{table-param_pic}. The Lorentz factor of the shock is fixed at $\gamma_{\rm sh}=1.86$. Depending on the value of $\theta_{B}$, the shock can be categorized as perpendicular (superluminal case), oblique (where extra energy is required for a particle to escape the shock), or parallel (where no extra energy is needed for a particle to escape the shock), with the critical transitions occurring at $\theta_B = 27.8$ (parallel-to-oblique) and $\theta_B = 32$ (oblique-to-perpendicular), measured in the upstream reference frame.

To study the shock evolution, we followed the simulations for a duration of $220\,\omci$, during which the electromagnetic shock precursor formed and the ion injection stabilized. The parameters of the stabilized ion injection are summarized in Table~\ref{table-param_pic}. As expected, the number density of shock-reflected particles remains stable for parallel shocks (PIC01-04). For oblique shocks (PIC05), the number density decreases, and no injection occurs in the superluminal case (PIC06). The normalized energy of injected particles, $U_{\rm inj}/U_0$, is found to be minimal for simulations PIC01-02, as particles do not require extra energy to outrun the shock in these cases. In contrast, simulations PIC03-04 show an increase in $U_{\rm inj}/U_0$, indicating that Shock Drift Acceleration (SDA) is active at these obliquities, despite the shock remaining parallel, and ions not requiring extra energy to escape the shock.

\begin{table*}[!t]
       \begin{centering}
\begin{tabular}{ccccccccccc}
\hline
\hline
\noalign{\smallskip}
Name  &  $\theta_B$ & $\theta_B$ & $N_{\rm inj}/N_{\rm 0}$ & $U_{\rm inj}/U_{\rm sh}$ &  $U_{\rm inj}/U_{\rm sh}$  \\
  &  SHRF & UFR &  &   URF &  SHRF  \\
\noalign{\smallskip}
\hline
\noalign{\smallskip}
PIC01  &  0    & 0    & 0.025 &  3.2  & 0.4 \\ 
PIC02  &  17.7 & 6.5  & 0.019 &  3.7  & 0.5   \\  
PIC03  &  33.4 & 13.3 & 0.021 &  7.3  & 1.8   \\  
PIC04  &  46.3 & 20.5 & 0.023 &  12.6 & 4.3   \\  
PIC05  &  56.7 & 28.6 & 0.013 &  10.9 & 4.6   \\ 
PIC06  &  61.1 & 32.9 & 0      &  -    & -     \\
\noalign{\smallskip}
\hline
\end{tabular}
\caption{PIC Simulation Parameters. $\theta_B$ is the obliquity angle measures in the upstream reference frame (URF) and the shock reference frame (SHRF). For the shock Lorentz factor of 1.86, the parallel-to-oblique transition at $\theta_B = 27.8$, the oblique-to-perpendicular transition at $\theta_B = 32$, where $\theta_B$ is measured in URF. $U_{\rm inj}/U_{\rm sh} = (\gamma_{\rm inj}-1)/(\gamma_{\rm sh}-1)$, where $\gamma_{\rm inj}$ is the average energy of injected particles in the respective reference frame and $(\gamma_{\rm sh}-1)mc^2$ is the energy of the upstream particles in the shock reference frame.}
\label{table-param_pic}
       \end{centering}
\end{table*}

\section{PIC-MHD simulation}

To investigate the large scale evolution of the shock, we perform a 2.5D PIC-MHD simulation, using a relativistic MHD version of the code described in \citep{vMCM18}, which is based on the {\tt MPI-AMRVAC} code \citep{VKM08}. This code combines aspects of both traditional {\bf MHD} and PIC in that it treats the thermal plasma as a fluid through MHD and the non-thermal charged particles through the PIC method. These two components interact with each other through the equation of motion for the particles and a modified version of Ohm's law for the fluid \citep{BCSS15}. We set up our simulations in the rest-frame of the shock (Fig.~\ref{fig:setup}(b)), starting from the analytical solution of the Rankine-Hugoniot conditions. Once the simulation has started, we begin to introduce the non-thermal particles at the shock front according to the injection rate given in Table~\ref{table-param_pic}.

For the PIC-MHD simulation, we start from a standing shock with a Lorentz factor of 1.86, and an obliquity of 30 degrees in the upstream rest-frame. At the shock, we inject non-thermal protons at a rate of 0.5 percent of the number of ions passing through the shock (hence compatible with PIC05 results). The non-thermal particles have a Lorentz factor of 3 times that of the pre-shock gas and are injected with an isotropic velocity distribution in the rest-frame of the shock.  The simulation space has a box-size of 180$\times$30\,$R_L$ with $R_L$ the Larmour radius defined by the velocity of the injected particles and the upstream magnetic field. In order to assure adequate resolution of the particle motion, we use a fixed grid with individual cells of 0.15\,$R_L$.

\begin{figure}
    \centering
    \includegraphics[width=0.49\linewidth]{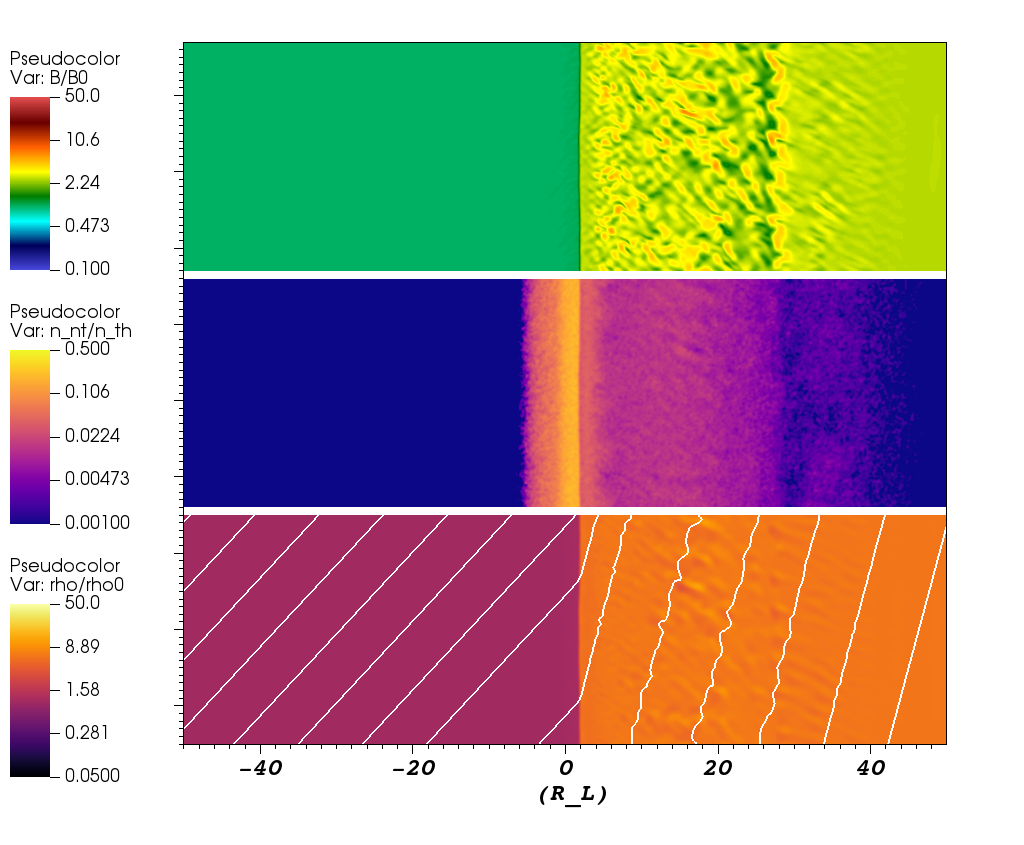}
    \includegraphics[width=0.49\linewidth]{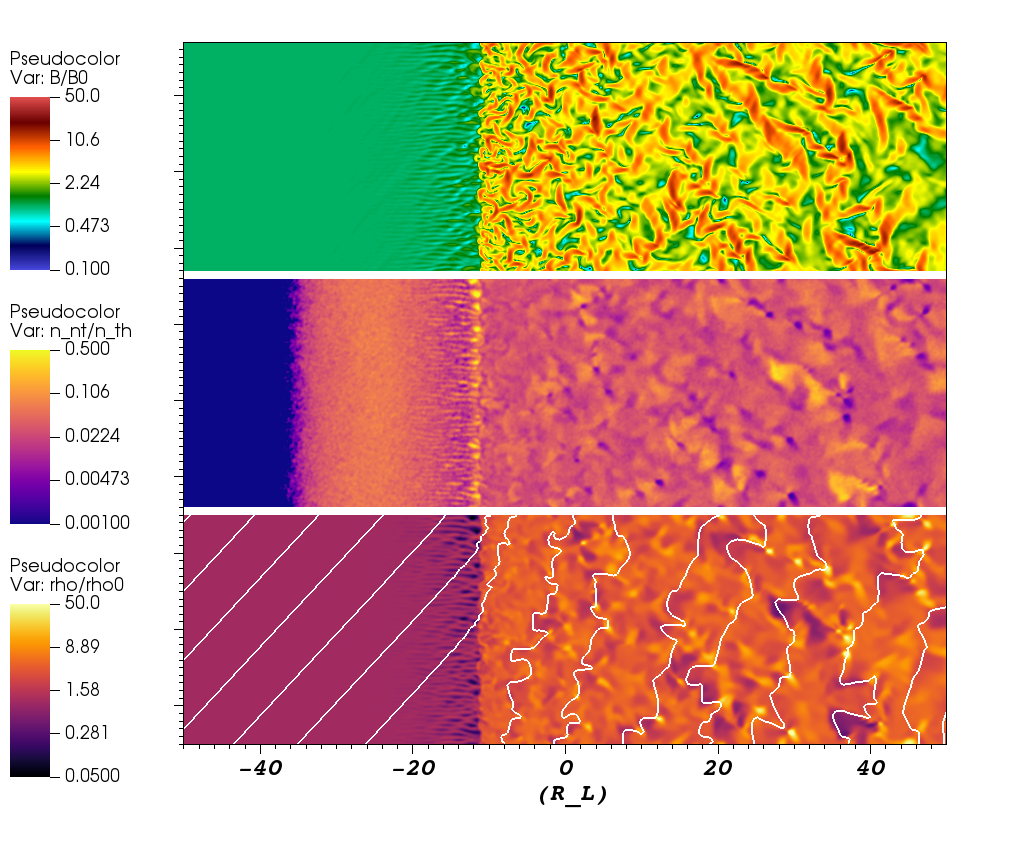}
    \caption{PIC-MHD results. From top to bottom, the figures show the magnetic field relative to the initial upstream field, the ratio of non-thermal to thermal particles, and the thermal gas density relative to the initial upstream density at $t\,=\,100 \ R_L/c$ (left) and $500 \ R_L/c$ (right). The bottom panel also includes the magnetic field lines.}
    \label{fig:picmhdshock}
\end{figure}

Figure~\ref{fig:picmhdshock} shows the result of the PIC-MHD simulation, displaying the magnetic field relative to the initial upstream field, the relative density of non-thermal to thermal particles, and the thermal gas density relative to the initial upstream density at $t\,=\,100$ and $500\,R_L/c$.
As can be seen, magnetic field shows both upstream and downstream distortion, which enables the reflection of non-thermal particles across the shock. However, the upstream disturbance is relatively small. 
As a result, we would expect that diffusive shock acceleration will be relatively inefficient, since many particles will escape  without being reflected. This situation is further enhanced by the size of the perturbed pre-shock region.

\begin{figure}
    \centering
    \includegraphics[width=0.70\linewidth]{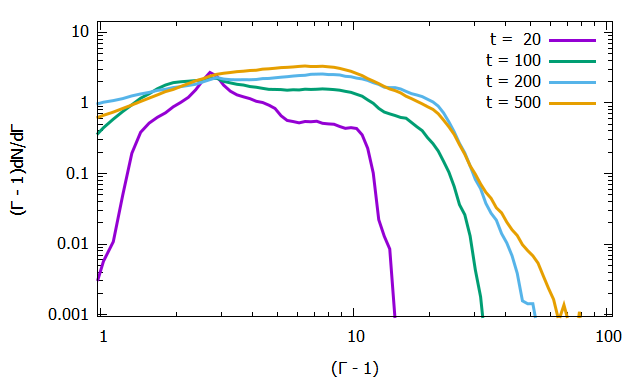}
    \caption{Spectral energy distribution over time for the PIC-MHD simulation.}
    \label{fig:picmhdsed}
\end{figure}

The spectral energy distributions {\bf resulting from the PIC-MHD simulation are }
shown in Fig,~\ref{fig:picmhdsed}. Initially, the spectrum is broad, but lacks an exponential high-energy tail, indicating that shock-drift is the primary acceleration mechanism. A high-energies the tail starts to appear over time, signalling that diffusive shock acceleration has started. However, a longer simulation is required in order to determine whether diffusive shock acceleration can become efficient in this particular situation.

\section{Conclusions}

\begin{itemize}
    \item Utilizing a fully-relativistic electromagnetic 2D3V PIC code, we perform a series of shock simulations with varying obliquity angles ($\theta_{B}$) to study the shock dynamics and particle acceleration processes. Stable shock reflected particle densities are observed for parallel shocks, whereas it decreases for oblique shocks and it  is absent in the superluminal case. Results are consistent with theoretical predictions for the selected Lorentz factor of the shock.

    \item The PIC-MHD simulation reveals that the magnetic field exhibits both upstream and downstream turbulence, enabling the reflection of non-thermal particles across the shock. However, the upstream disturbance is relatively small, leading to the expectation that diffusive shock acceleration will be relatively inefficient and longer simulations are required to determine whether diffusive shock acceleration will become efficient in this specific scenario.
\end{itemize}

\section*{Acknowledgments}
A.~B. was supported by the German Research Foundation (DFG) as part of the Excellence Strategy of the federal and state governments - EXC 2094 - 390783311. Computations were performed on the HPC system Raven at the Max Planck Computing and Data Facility.

\end{document}